\documentclass[reprint,aps,prc,showpacs,superscriptaddress,preprintnumbers,amsmath,amssymb,amsfonts]{revtex4-2}
\usepackage{graphicx,latexsym,amssymb,amsmath,color,multirow,mathrsfs,booktabs,ifpdf}
\def\bm#1{\mbox{\boldmath $#1$}}
\def\nA{nucleon-nucleus\ }
\def\aA{$\alpha$-nucleus\ }
\def\oc{$^{16}$O+$^{12}$C\ }

\def\cc{$^{12}$C+$^{12}$C\ }
\def\oo{$^{16}$O+$^{16}$O\ }
\def\AA{nucleus-nucleus\ }
\def\Atrans{$^{12}$C($^{16}$O,$^{12}$C)$^{16}$O\ }
\begin{document}
\title{Nuclear rainbow of the symmetric nucleus-nucleus system: 
 \\ Interchange of the nearside and farside scattering}

\author{Nguyen Tri Toan Phuc}
\affiliation{Department of Nuclear Physics, Faculty of Physics and Engineering Physics, University of Science, 
Ho Chi Minh City 700000, Vietnam}
\affiliation{Vietnam National University, Ho Chi Minh City 700000, Vietnam}

\author{Nguyen Hoang Phuc}
\email{phuc.nguyenhoang@phenikaa-uni.edu.vn}
\affiliation{Phenikaa Institute for Advanced Study (PIAS), Phenikaa University, Hanoi 12116, Vietnam}

\author{Dao T. Khoa}
\affiliation{Institute for Nuclear Science and Technology, VINATOM, Hanoi 122772, Vietnam}
\begin{abstract}
Extensive elastic scattering data measured at energies around 10 to 20 MeV/nucleon for some 
light identical systems, like \cc and \oo, were shown to exhibit the nuclear rainbow pattern 
of broad Airy oscillations of the elastic scattering cross section at medium and large angles. 
Because of the identity of the scattered projectile and recoiled target, the smooth rainbow pattern 
at angles around and beyond $\theta_{\rm c.m.}\approx 90^\circ$ is strongly deteriorated by the 
boson exchange in the \cc and \oo systems at low energies. The exchange symmetry of two identical 
nuclei implies the Mott interference of the direct and exchange scattering amplitudes, 
which destroys the nuclear rainbow pattern. 
The nuclear rainbow features in the elastic scattering of two identical nuclei have been 
discussed so far based on the nearside-farside (NF) decomposition of the scattering amplitude 
given by an optical model calculation neglecting the projectile-target exchange symmetry. 
Moreover, the NF decomposition method was developed in the 1970s by Fuller for 
\emph{nonidentical} dinuclear systems only, and the details of how the exchange symmetry 
of an \emph{identical} system affects the evolution of nuclear rainbow remain unexplored. 
For this purpose, the Fuller method is generalized in the present work for the elastic scattering 
of two identical (spin-zero) nuclei, with the projectile-target exchange symmetry taken explicitly 
into account. The results obtained for elastic \cc and \oo scattering at low energies show that 
the exchange symmetry results in a symmetric interchange of the nearside and farside scattering 
patterns at angles passing through $\theta_{\rm c.m.}=90^\circ$, which requires a more subtle 
interpretation of nuclear rainbow. 
We found further that a similar NF interchange also occurs in a nonidentical \AA system with 
the core-core symmetry at low energies, where the elastic cross section at backward angles is due 
mainly to the elastic transfer of cluster or nucleon between two identical cores. This interesting 
effect is illustrated in the elastic \oc scattering at low energies where the elastic $\alpha$ transfer 
between two $^{12}$C cores has been proven to enhance the elastic cross section at backward angles.
\end{abstract}
\maketitle

\section{Introduction}\label{sec1} 
\begin{figure}[bt]\vspace*{0cm}\hspace*{0cm}
	\includegraphics[angle=0,width=0.45\textwidth]{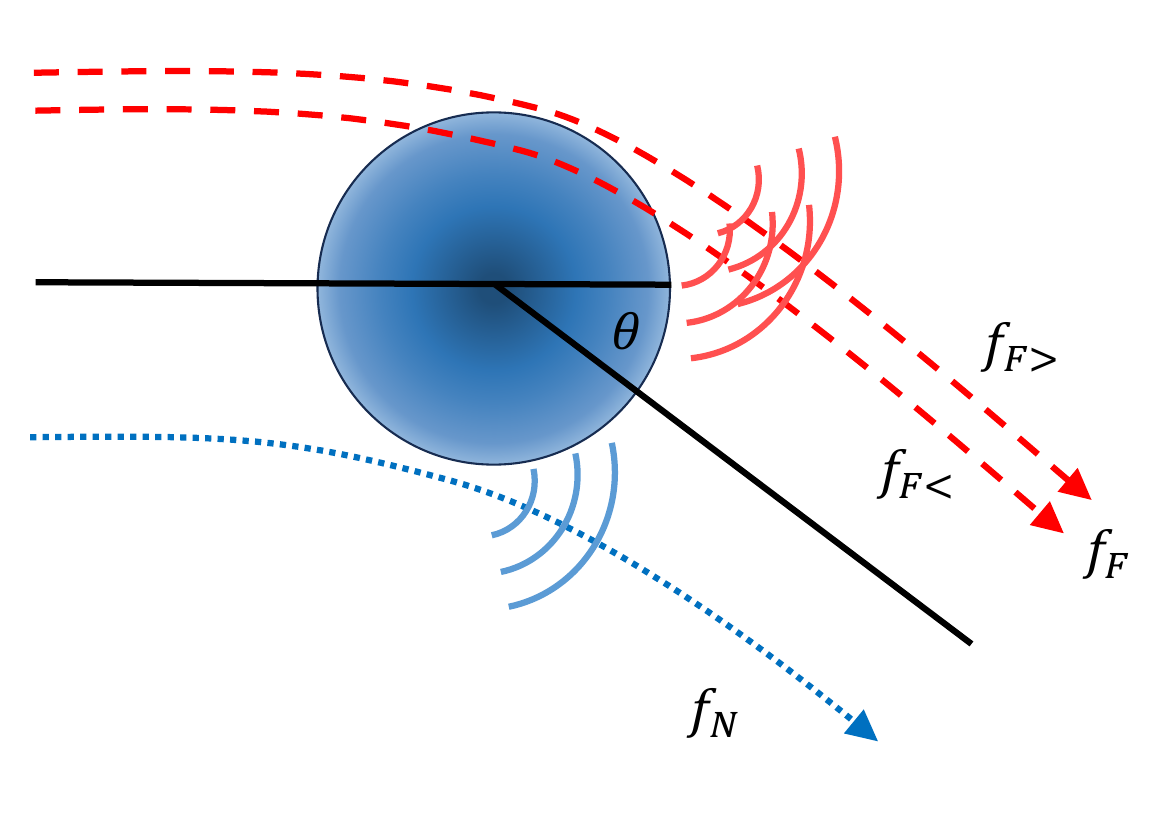}\vspace*{0cm}
	\caption{Schematical view of trajectories of the incident wave scattered by nuclear 
		OP at the surface and sub-surface impact parameters, to the nearside and farside
		of the scattering center, respectively. $f_{F>}$ and $f_{F<}$ are the farside subamplitudes 
		with $\ell>\ell_{\rm R}$ and $\ell<\ell_{\rm R}$, respectively, where $\ell_{\rm R}$ is the 
		angular momentum at the rainbow point (see discussion in text).} \label{f1}
\end{figure}

Elastic heavy-ion (HI) scattering is usually associated with strong absorption, 
with the scattering cross section displaying a typical diffraction pattern at forward angles 
and falling down rapidly at larger angles \cite{Sa79}. Such a diffraction pattern 
results mainly from the surface scattering of the incident wave, and it can be described 
by different choices of the optical potential (OP), illustrating the well-known ambiguity
of the HI optical potential \cite{Sa79,Bra97}. However, many $\alpha$-nucleus and several 
light HI systems were found to be quite weakly absorbing, enabling the (refractive) 
nuclear rainbow pattern to appear at medium and large scattering angles \cite{Bra97,Kho07r}. 
Such a rainbow pattern originates from the refraction of incident wave at smaller (sub-surface) 
impact parameters that shows up in the scattering cross section at larger angles if the 
absorption is weak (see Fig.~\ref{f1}). As a consequence, the observation of nuclear rainbow 
allows us to probe the \AA interaction at both the surface and sub-surface distances, and 
to determine the real \AA OP with much less ambiguity (see, e.g., the topical review \cite{Kho07r} 
for more details). 

To motivate the present study, we briefly recall the main features of the nuclear rainbow 
phenomenon. When the absorption is weak, the interference pattern of scattering waves 
attractively refracted by the target (serving as a nuclear liquid drop) turns out to be analogous 
to the interference of light rays refracted by a water droplet that causes the atmospheric rainbow. 
As schematically illustrated in Fig.~\ref{f1}, the trajectories incident at the sub-surface 
impact parameters are attractively refracted to the \emph{far side} of the scattering center 
\cite{Nuss77,Hus84}, and the interference of the two subamplitudes of the farside scattering 
($f_{F>}$ and $f_{F<}$)  gives rise to a broad oscillation pattern of elastic scattering cross
section, which is a microscopic replica of the Airy oscillation pattern of the refracted light rays 
of atmospheric rainbow \cite{Kho07r,Bra96}. Thus, the key to identify a nuclear rainbow is the 
Airy oscillation of the farside cross section in elastic \AA scattering, especially, the first Airy 
minimum A1 followed by a shoulder-like bump in elastic scattering cross section at medium and 
large angles \cite{Kho07r,Bra96,Fri88}. 

The farside scattering pattern can be revealed by decomposing the elastic scattering amplitude 
into the \emph{internal} component that penetrates the Coulomb + centrifugal barrier into the 
interior of the real OP, and the \emph{barrier} component that is reflected from the barrier 
\cite{Bri77,Row77}. 

A more popular, alternative interpretation of the farside scattering is based on the decomposition 
of the elastic scattering amplitude into the \emph{nearside} and \emph{farside} components, using
the method originally developed by Fuller  \cite{Ful75} which is referred to hereafter as the 
nearside-farside (NF) decomposition. The Airy oscillation pattern of the nuclear rainbow is formed 
by an interference of the outer ($f_{F>}$) and inner ($f_{F<}$) subamplitudes of the farside scattering. 
The $f_{F>}$ and $f_{F<}$ subamplitudes represent the contributions to the same scattering angle 
$\theta$ of the (farside) partial waves with angular momenta $\ell>\ell_{\rm R}$ and $\ell<\ell_{\rm R}$, 
respectively, where $\ell_{\rm R}$ is the angular momentum at the minimum of the deflection function, the 
\emph{rainbow point} \cite{McVoy86} that determines the corresponding rainbow angle $\Theta_{\rm R}$ 
in the observable angular range ($0<\Theta_{\rm R}<180^\circ$) \cite{FroLip96}. The scattering 
to angles $\theta>\Theta_{\rm R}$,  i.e., to the dark side of the rainbow, is classically forbidden. 
The location of $\Theta_{\rm R}$ is shifting from large scattering angles at low energies to medium 
angles of $\Theta_{\rm R}\sim 40^\circ-60^\circ$ at energies approaching Fermi domain 
of around 20 MeV/nucleon, where pronounced nuclear rainbow patterns have been observed 
\cite{Kho07r}. It is complementary to note that $\Theta_{\rm R}\approx 138^\circ$ for the refracted 
light rays of atmospheric rainbow \cite{Nuss77}. In elastic HI scattering, the $f_{F<}$ subamplitude 
of the farside scattering is often suppressed by the absorption, and the Airy oscillation pattern of nuclear 
rainbow disappears. Therefore, a weak absorption is a prerequisite for the formation and observation 
of nuclear rainbow.      

As mentioned above, the observation of the (refractive) nuclear rainbow pattern provides an
important database for studies of the \AA interaction at low and medium energies \cite{Bra97,Kho07r}. 
Because a nuclear rainbow is formed mainly by the farside trajectories refracted by a 
strongly attractive \AA OP, the nuclear rainbow pattern is absent in elastic electron-nucleus  
or \nA scattering where the scattering potential is not attractive enough to give rise to the
farside scattering \cite{Bra96}. Thus, the nuclear rainbow can appear only in the refractive 
\aA or \AA scattering which is governed by the deep attractive real OP \cite{Kho07r}. 

The nuclear rainbow pattern has been shown mainly so far by using the Fuller decomposition
method \cite{Ful75}. However, this method was formulated for a \emph{nonidentical} 
dinuclear system only, and it remains unexplored how the projectile-target exchange 
symmetry of an \emph{identical} system affects the nuclear rainbow pattern. For this purpose, 
we have generalized in the present work, the Fuller method for the elastic scattering of two identical 
(even-even) nuclei, taking into account exactly the projectile-target exchange symmetry. The generalized 
method is then used to study the nuclear rainbow pattern in elastic \cc and \oo scattering, which was
proven to be strongly refractive \cite{Bra88,Mcv92,Kho94,Kho97,Kho00,Kho07r,Kho16}. 

\begin{figure}[bt]\vspace*{0cm}\hspace*{-0.5cm}
	\includegraphics[width=0.55\textwidth]{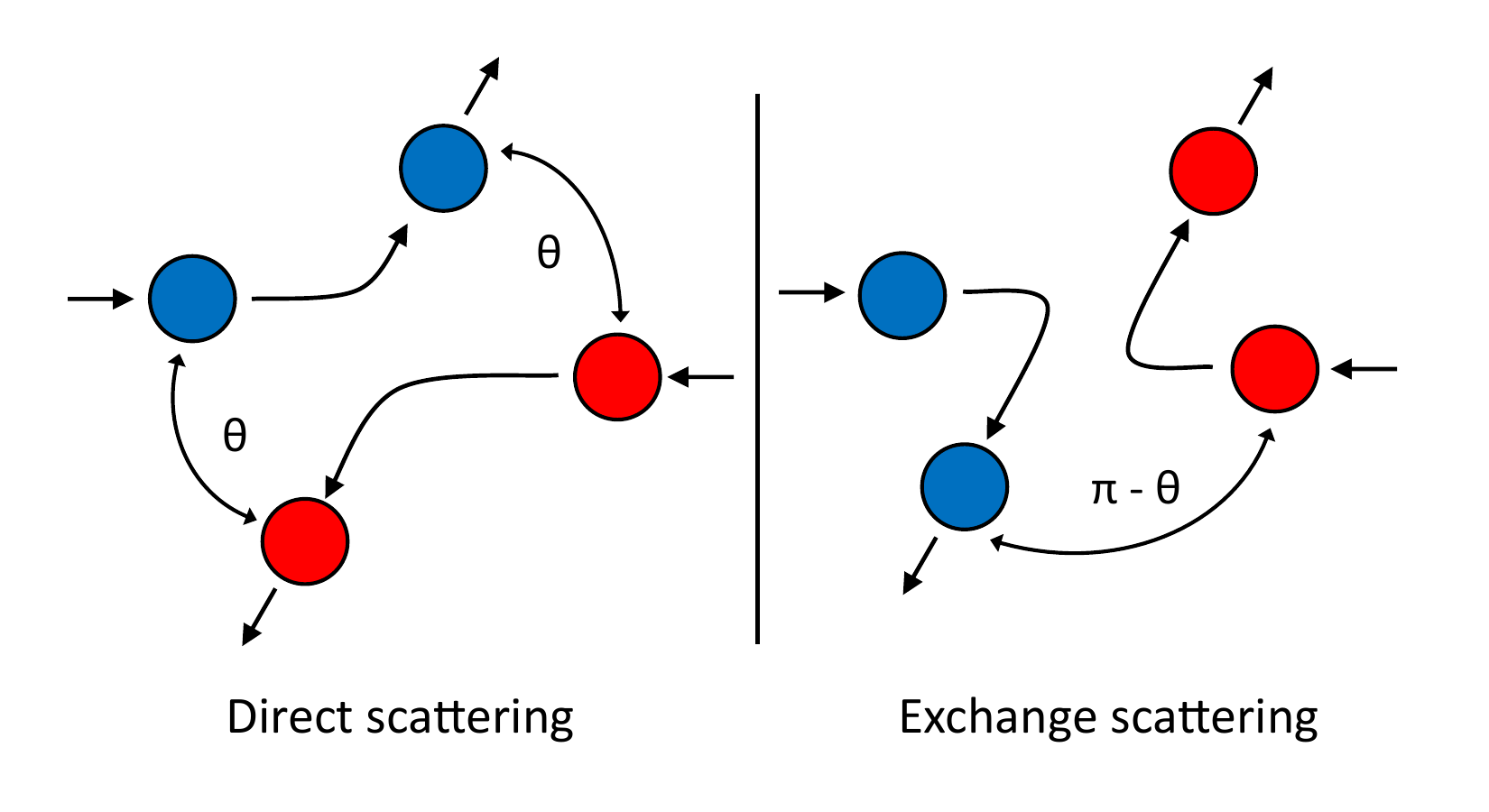}\vspace*{0cm}
	\caption{Kinematical illustration of the direct and exchange scattering of two
		identical nuclei.} \label{f2}
\end{figure}
Because of the identity of two nuclei, the detector cannot distinguish between the scattered 
projectile (in the direct scattering) and recoiled target (in the exchange scattering) as illustrated 
in Fig.~\ref{f2}, and the total scattering wave function of an identical \AA system must be 
composed of both the direct and exchange components. Then, the total wave function becomes 
symmetric or antisymmetric with respect to the projectile-target exchange \cite{Sat83} for 
the integer or half-integer nuclear spin, respectively, 
\begin{equation}
\Psi_{\rm total}({\bm r})\sim\left[\Psi({\bm r})\pm\Psi(-{\bm r})\right],\ 
{\bm r}={\bm r}_1-{\bm r}_2. \label{eq1}
\end{equation}
As a result, the total elastic scattering (ES) amplitude is also symmetric or antisymmetric in the same 
manner, with respect to the projectile-target exchange \cite{FroLip96} 
\begin{equation}
f_{\rm ES}(\theta)=f_{\rm Mott}(\theta)+f(\theta)\pm f(\pi-\theta), \  \  \theta\equiv\theta_{\rm c.m.}. 
\label{eq2}
\end{equation}
where $f_{\rm Mott}(\theta)$ is the \emph{Coulomb} scattering amplitude of two identical ions, known as the 
Mott scattering amplitude \cite{Sat83}, $f(\theta)$ and $f(\pi-\theta)$ are the direct and  exchange 
\emph{nuclear} scattering amplitudes. The symmetrization or antisymmetrization procedure (\ref{eq2}) 
is commonly available in different OM or coupled-channel codes  \cite{Raynal,Tho88,Tho09}. 

\begin{figure}[bt]\vspace*{0cm}\hspace*{-0.5cm}
\includegraphics[width=0.54\textwidth]{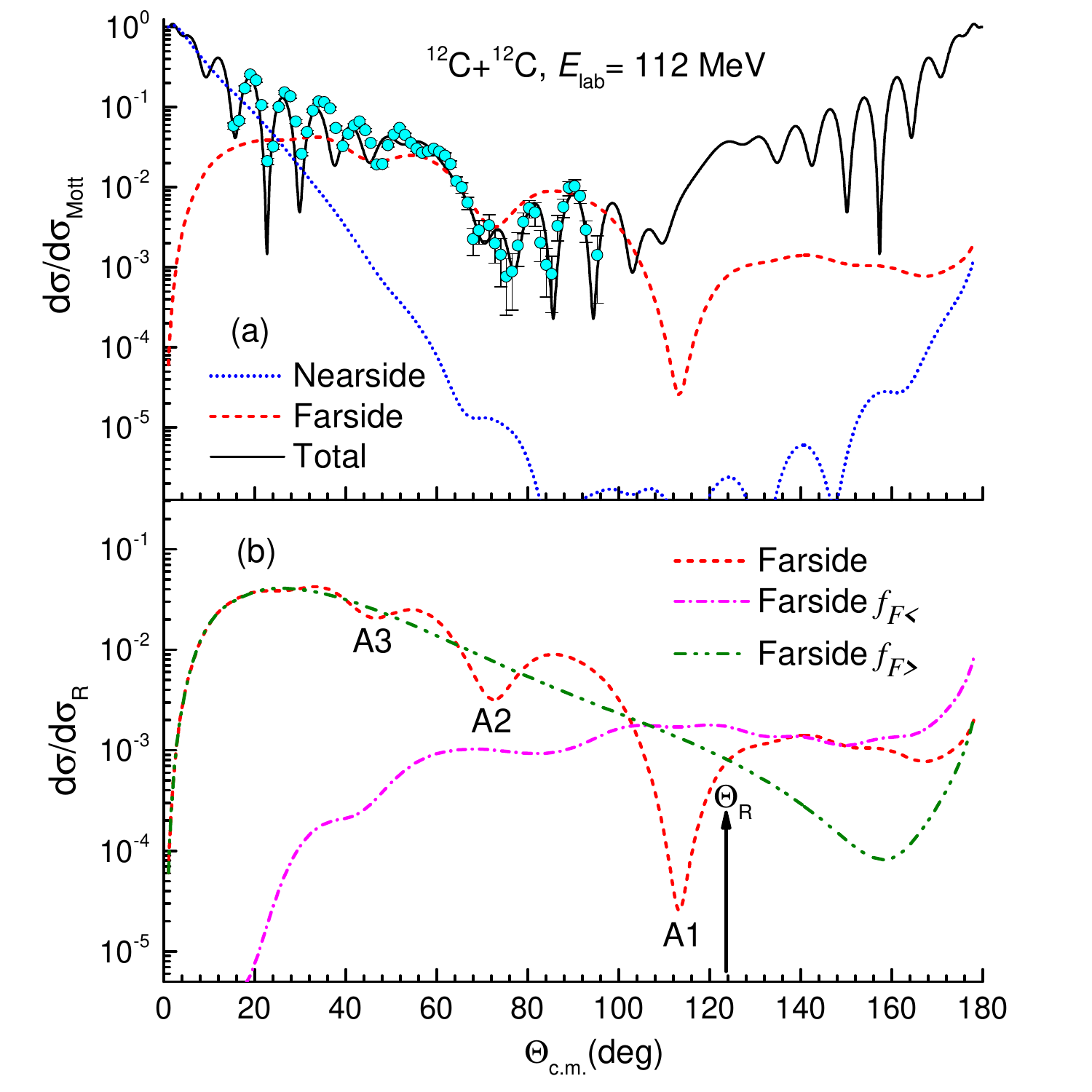}\vspace*{0cm}
 \caption{(a) Elastic \cc scattering data measured at $E_{\rm lab}=112$ MeV 
\cite{Sto79} in comparison with the OM results, taking explicitly into account the boson 
symmetrization (\ref{eq1})-(\ref{eq2}) of two identical $^{12}$C nuclei (solid line). The nearside 
(dotted line) and farside (dashed line) cross sections were obtained from the NF decomposition 
(\ref{eq3})-(\ref{eq6}) of the elastic scattering amplitude, neglecting the projectile-target symmetry. \\
(b) The full farside cross section (dashed line) and cross sections given by $f_{F>}$ (dashed dotted 
dotted line) and $f_{F<}$ (dashed dotted line) farside subamplitudes. Ak is the $k$-order Airy 
minimum, and $\Theta_{\rm R}$ is the rainbow angle determined from the minimum of the deflection 
function \cite{FroLip96}.} \label{f3}
\end{figure}
A typical example is the elastic \cc scattering data measured at $E_{\rm lab}=112$ MeV shown in 
Fig.~\ref{f3}, which can be properly described only when the projectile-target exchange symmetry
(\ref{eq2}) is taken into account in the OM calculation. So far, the nuclear rainbow pattern in elastic 
\cc scattering has been studied only qualitatively with the projectile-target exchange symmetry (\ref{eq2}) 
turned off in the OM calculation, and the standard (unsymmetrized) NF decomposition \cite{Ful75} 
of the elastic scattering amplitude could be made to reveal the Airy oscillation of the farside cross section. 
One can see in panel (b) of Fig.~\ref{f3} that such an Airy oscillation is given by an interference of the two 
subamplitudes of the farside scattering. The outer $f_{F>}(\theta)$ and inner $f_{F<}(\theta)$ subamplitudes 
were determined in this work based on the method suggested by McVoy {\it et al.} \cite{McVoy86}. 
However, the broad Airy oscillation of the (unsymmetrized) farside cross section from the rainbow angle 
$\Theta_{\rm R}\approx 123^\circ$, over A1 and A2 to medium angles, is strongly distorted when 
the projectile-target symmetrization is taken explicitly into account as shown in panel (a) of Fig.~\ref{f3}, 
and the elastic scattering cross section becomes symmetric with respect to the angular exchange 
$\theta \leftrightarrow \pi-\theta$. It is also obvious from the exchange symmetry that the quickly 
oscillating pattern at the most backward angles is due to a NF interference similar to that established 
at forward angles, but the original Fuller method \cite{Ful75} is not appropriate for that kind of NF 
analysis. In fact, a proper determination of the nearside and farside scattering amplitudes 
for a symmetric \AA system remains an unsolved problem. 
  
\section{NF decomposition of the elastic scattering amplitude of two identical spin-zero nuclei}
\label{sec2}
We briefly recall how the nuclear rainbow pattern is revealed by decomposing the elastic scattering 
amplitude into the nearside $f^{(\rm N)}$ and farside $f^{(\rm F)}$ components using 
Fuller method \cite{Ful75}. By splitting the Legendre function $P_\ell(\cos\theta)$ into two waves 
scattered at $\theta$ but running in the opposite directions around the scattering center, the NF 
components of the unsymmetrized (nuclear) elastic scattering amplitude is obtained as 
\begin{subequations} \label{eq3}
\begin{eqnarray}
  f(\theta)&=&f^{(\rm N)}(\theta)+f^{(\rm F)}(\theta),  \label{eq3a} \\
	f^{(\rm N)}(\theta)&=&\frac{i}{2k}\sum_\ell (2\ell+1)B_\ell\tilde Q_\ell^{(-)}(\cos\theta), \label{eq3b} \\
	f^{(\rm F)}(\theta) &=&\frac{i}{2k}\sum_\ell (2\ell+1)B_\ell\tilde Q_\ell^{(+)}(\cos\theta). \label{eq3c}
\end{eqnarray}
\end{subequations}
The partial-wave amplitude $B_\ell$ is determined as
\begin{equation} 
B_\ell = \exp(2i\sigma_\ell)(1-S_\ell),
\end{equation}
where $\sigma_\ell$ is the Coulomb phase shift, and $S_\ell$ is the scattering $S$ matrix element 
for the $\ell$-th partial wave. The traveling wave components are
\begin{equation}
 \tilde Q_\ell^{(\mp)}(\cos\theta)={1\over 2}
 \left[P_\ell(\cos\theta)\pm {2i\over\pi}Q_\ell(\cos\theta)\right],
\end{equation}
where $Q_\ell(\cos\theta)$ is the Legendre function of the second kind. A nice feature of the quantal 
scattering theory is that the wave associated with $Q_\ell^{(-)}(\cos\theta)$ is deflected from the 
\emph{near side} of the scattering center to angle $\theta$, and that associated with 
$Q_\ell^{(+)}(\cos\theta)$ is deflected from the opposite, \emph{far side} of the scattering center 
to the same angle $\theta$. Thus, the nearside amplitude $f^{(\rm N)}(\theta)$ accounts mainly 
for the \emph{repulsively} diffracted wave that scatters at the surface, and the farside amplitude 
$f^{(\rm F)}(\theta)$ accounts for the \emph{attractively} refracted wave that penetrates more 
into the sub-surface of the target nucleus, as schematically shown in Fig.~\ref{f1}. 

Because of the short range of the strong interaction, the NF decomposition of the \emph{nuclear} 
scattering amplitude (\ref{eq3}) can be done numerically, with the nuclear nearside and farside
amplitudes merging naturally at $\theta=0$ and $\theta=\pi$. However, the long-range Coulomb 
interaction poses a technical difficulty for the NF decomposition of the \emph{Coulomb} 
scattering amplitude based on the partial wave expansion (\ref{eq3}). Moreover, the Coulomb 
scattering amplitude is singular at $\theta=0$, where the partial wave series diverges. 
By projecting argument of the Legendre function into the complex plane, Fuller managed to treat 
the singularity at $\theta=0$ and obtained the NF components $f^{\rm (N)}_{\rm R}(\theta)$  
and $f^{\rm (F)}_{\rm R}(\theta)$ of the Rutherford scattering amplitude  \cite{Ful75} as 
\begin{subequations} \label{eq6}
\begin{align}
f^{\rm (N)}_{\rm R}(\theta)=f_{\rm R}(\theta)\left\{\frac{-iS(\theta)}{2\pi}
 \left[\sin^2(\theta/2)\right]^{1+i\eta}+ {1-e^{2\pi\eta}}\right\},  \hskip 0.6cm\label{eq6a} \\ 
f^{\rm (F)}_{\rm R}(\theta)=f_{\rm R}(\theta)\left\lbrace\frac{iS(\theta)}{2\pi}
 \left[\sin^2(\theta/2)\right]^{1+i\eta}- \frac{e^{-2\pi\eta}}{1-e^{2\pi\eta}}\right\rbrace. 
\hskip 0.8cm \label{eq6b}\\
f_{\rm R}(\theta)=-\dfrac{\eta}{2ik \sin^2(\theta/2)}
\exp\left[2i\sigma_0-i\eta \ln\sin^2(\theta/2)\right], \hskip 1cm\label{eq6c}\\
S(\theta)=(1+i\eta)^{-1}F(1,1+i\eta;2+i\eta;\sin^2(\theta/2)), \hskip 1.3cm \\ 
\mbox{and hypergeometric function}\ \hskip 1.8cm \nonumber \\ 
F(a,b;c;z)={ }_{2}F_{1} (a,b;c;z), \hskip 4.5cm \nonumber \\ 
=\frac{\Gamma(c)}{\Gamma(a)\Gamma(b)}\sum_{n=0}^\infty 
\frac{\Gamma(a+n)\Gamma(b+n)}{\Gamma(c+n)}\frac{z^n}{z!}.\hskip 1.25cm\label{eq6d}
\end{align}
\end{subequations}
Here $k$ and $\eta$ are the wave number and Sommerfeld parameter, respectively,
and $\sigma_0=\arg\Gamma(1+i\eta)$ \cite{Sat83}. 

As discussed above, the elastic scattering cross section of two identical (spin-zero) 
nuclei is symmetric about $\theta=\pi/2$. In particular, the Coulomb scattering amplitude 
of two identical ions is known as the Mott scattering amplitude \cite{Sat83}, which is 
determined in this case from the Rutherford scattering amplitude (\ref{eq6b}) as
\begin{equation}
f_\text{Mott}(\theta)=f_\text{R}(\theta)+f_\text{R}(\pi-\theta). \label{eq7}
\end{equation}
After some trigonometric transformation, we obtain 
\begin{equation}
\hskip -0.3cm	f_{\rm Mott}(\theta)=f_R(\pi)\left[\left(\frac{1-x}{2}\right)^{-1-i\eta}+
	\left(\frac{1+x}{2}\right)^{-1-i\eta}\right], \label{eq8}
\end{equation}
where $x=\cos\theta$. Applying the Fuller technique, we have projected the Mott scattering 
amplitude (\ref{eq7})-(\ref{eq8})  into the complex plane, also in terms of the two components 
(in the forward and backward scattering angles \cite{Ful75}). After a lengthy analytical transformation, 
the nearside and farside components of the Mott scattering amplitude can be rigorously expressed 
in terms of those of the Rutherford amplitude (\ref{eq6a})-(\ref{eq6b}) as 
\begin{eqnarray}
f^{\rm (N)}_{\rm Mott}(\theta)&=&f^{\rm (N)}_{\rm R}(\theta)+f^{\rm (F)}_{\rm R}(\pi-\theta), 
 \label{eq9} \\
f^{\rm (F)}_{\rm Mott}(\theta)&=&f^{\rm (F)}_{\rm R}(\theta)+f^{\rm (N)}_{\rm R}(\pi-\theta). 
 \label{eq10}
\end{eqnarray}
Thus, the nearside component (\ref{eq9}) of the Mott amplitude is a superposition of the nearside 
and farside components of the Rutherford amplitude determined at the angles $\theta$ and $\pi-\theta$, 
respectively. Similarly, the farside component (\ref{eq10}) is a superposition of the farside and nearside 
components of the Rutherford amplitude determined at the angles $\theta$ and $\pi-\theta$, respectively. 
$f^{\rm (N)}_{\rm Mott}(\theta)$ and $f^{\rm (F)}_{\rm Mott}(\theta)$ become equal at $\theta=\pi/2$. 

The symmetrization of the nuclear scattering amplitude is straightforwardly based on the partial 
wave series (\ref{eq3}), and the symmetrized amplitude can be decomposed into the nearside 
and farside components as
\begin{align} 
f_{\rm sym}(\theta)=f^{\rm (N)}_{\rm sym}(\theta)+f^{\rm (F)}_{\rm sym}(\theta)
\hskip 3.8cm \nonumber\\
 =\frac{1}{2ik}\sum_\ell B_\ell\left[1+(-1)^\ell\right] 
\left[\tilde Q_\ell^{(-)}(\cos\theta) +\tilde Q_\ell^{(+)}(\cos\theta)\right], \label{eq11}
\end{align}
where the summation is done over the \emph{even} partial waves $\ell$ only.
Given $(-1)^\ell\tilde Q_\ell^{(\pm)}(\cos\theta)=\tilde Q_\ell^{(\mp)}(\cos(\pi-\theta))$,
the nearside and farside components of the symmetrized nuclear amplitude (\ref{eq11}) are 
readily obtained as
\begin{eqnarray} 
f^{\rm (N)}_{\rm sym}(\theta)&=&f^{(\rm N)}(\theta)+f^{(\rm F)}(\pi-\theta),  \label{eq12} \\
f^{\rm (F)}_{\rm sym}(\theta)&=& f^{(\rm F)}(\theta)+f^{(\rm N)}(\pi-\theta),  \label{eq13}
\end{eqnarray} 
where $f^{(\rm N,F)}$ are the nearside and farside components of the unsymmetrized
nuclear scattering amplitude (\ref{eq3b})-(\ref{eq3c}). Thus, exactly in the same way as 
for the Mott scattering amplitude, the nearside component of the symmetrized nuclear 
scattering amplitude (\ref{eq11}) is also a superposition of the nearside and farside components 
of the unsymmetrized nuclear amplitude (\ref{eq3a}) summed over the even partial waves $\ell$ 
at the angles $\theta$ and $\pi-\theta$, respectively, and vice versa for the farside component 
(\ref{eq11}). 

It is helpful to express explicitly the total ES amplitude of two identical (spin zero) nuclei 
in terms of the direct (D) and exchange (EX) scattering amplitudes, as illustrated in Fig.~\ref{f2},
\begin{equation}
	f_{\rm ES}(\theta)=f_{\rm D}(\theta)+f_{\rm EX}(\pi-\theta), \label{eq14}
\end{equation}
where 
\begin{subequations} \label{eq15}
\begin{eqnarray} 
f_{\rm D}(\theta)&=&f_{\rm R}(\theta)+f(\theta),   \\  
f_{\rm EX}(\pi-\theta)&=&f_{\rm R}(\pi-\theta)+f(\pi-\theta).
\end{eqnarray} 
\end{subequations}
It is easy to deduce from Eqs.~(\ref{eq14})-(\ref{eq15}) the symmetric interchange of the direct 
and exchange scattering amplitudes $f_{\rm D}\rightleftarrows f_{\rm EX}$ with the scattering 
angle passing through $\theta=90^\circ$.
Combining the nearside and farside components of the Mott (\ref{eq9})-(\ref{eq10}) and 
nuclear (\ref{eq12})-(\ref{eq13}) scattering amplitudes, we can express the nearside and 
farside components of the ES amplitude as 
\begin{eqnarray}
f^{\rm (N)}_{\rm ES}(\theta)&=&f^{\rm (N)}_{\rm D}(\theta)+
f^{\rm (F)}_{\rm EX}(\pi-\theta), \label{eq16} \\
f^{\rm (F)}_{\rm ES}(\theta)&=&f^{\rm (F)}_{\rm D}(\theta)+
f^{\rm (N)}_{\rm EX}(\pi-\theta). \label{eq17} 
\end{eqnarray}
In the same way, one can deduce from Eqs.~(\ref{eq16})-(\ref{eq17}) the symmetric 
interchange of the \emph{nearside} and \emph{farside} components of the total ES amplitude 
$f^{\rm (N)}_{\rm ES}\rightleftarrows f^{\rm (F)}_{\rm ES}$ as the scattering angle 
passes through $90^\circ$. Because the direct NF cross sections become negligible  
at angles $\theta>90^\circ$ (see Fig.~\ref{f3}), we obtain from 
Eqs.~(\ref{eq16})-(\ref{eq17}) 
\begin{eqnarray} 
f^{\rm (N)}_{\rm ES}(\theta)&\approx&f^{\rm (F)}_{\rm EX}(\pi-\theta)
 \ {\rm at}\  \theta> 90^\circ,  \label{eq18} \\
f^{\rm (F)}_{\rm ES}(\theta)&\approx&f^{\rm (N)}_{\rm EX}(\pi-\theta) 
 \ {\rm at}\  \theta> 90^\circ. \label{eq19} 
\end{eqnarray}
Thus, the symmetric interchange of the nearside and farside scattering patterns in the elastic 
scattering of two identical nuclei is caused naturally by the projectile-target exchange symmetry. 
As shown in the next section, such a NF interchange implies a more subtle interpretation 
of nuclear rainbow over the whole angular range for the considered symmetric systems.

\section{Results for the elastic \cc and \oo scattering}\label{sec3} 
We present here the results of the NF decomposition of elastic \cc and \oo scattering 
at low energies, where the data were measured accurately up to angles around and beyond 
$\theta\approx 90^\circ$. The real OP is given by the double-folding calculation \cite{Kho16} 
using the CDM3Y3 density-dependent interaction \cite{Kho97}, and the imaginary OP is 
parametrized in the Woods-Saxon (WS) form to tailor the weak absorption of these systems. 
A slight renormalization $N_{\rm R}$ of the real folded potential and WS parameters are 
obtained from the best OM fit to the measured data using the code ECIS97 \cite{Raynal}. 
Very prominent are the \cc scattering data measured by Stokstad \emph{et al.} \cite{Sto79} 
at low energies, which cover a wide angular range. Based on these data, a realistic 
OM description of the nuclear rainbow pattern and $90^\circ$ excitation function was 
obtained by McVoy and Brandan \cite{Mcv92}, with the angular location of Airy minima  
at different energies (the famous Airy elephant discussed in Ref.~\cite{Mcv92}).

\begin{figure}[bt]\vspace*{0cm}\hspace*{-0.5cm}
\includegraphics[width=0.55\textwidth]{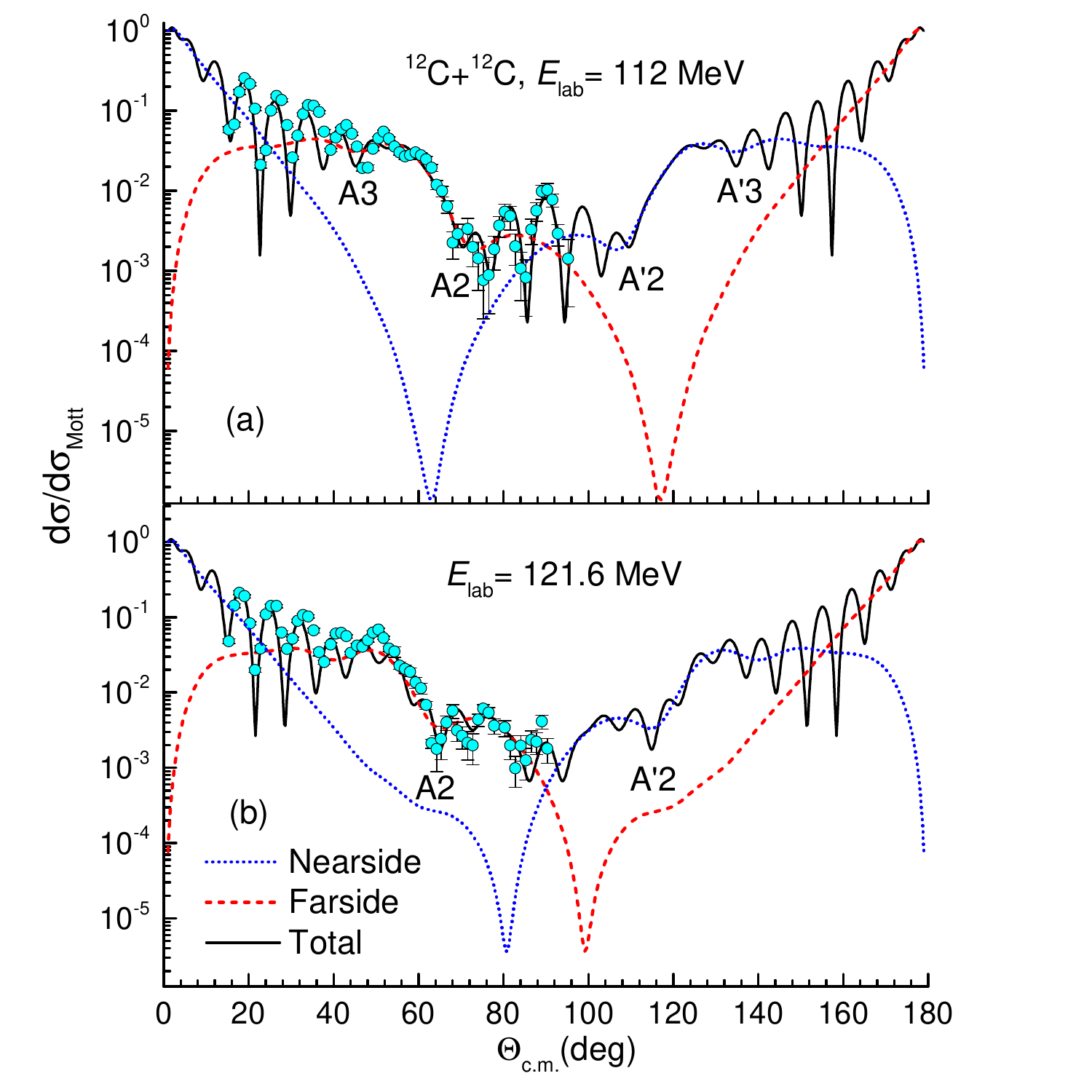}\vspace*{0cm}
 \caption{(a) Elastic \cc scattering data measured at $E_{\rm lab}=112$ MeV \cite{Sto79} 
in comparison with results of the OM calculation, taking exactly into account the 
projectile-target symmetrization (solid line). The nearside (dotted line) and farside (dashed line) 
cross sections were given by the NF decomposition (\ref{eq16})-(\ref{eq17}) of the ES amplitude. 
A$k$ and A'$k$ are the $k$-order Airy minimum of the direct farside cross section at $\theta$ and its 
symmetric partner of the exchange farside cross section at $\pi-\theta$, respectively. \\
(b) The same as (a) but for the elastic \cc scattering data measured at $E_{\rm lab}=121.6$ 
MeV \cite{Sto79}.} \label{f4}
\end{figure}
The OM results for elastic \cc scattering at $E_{\rm lab}=112$ and 121.6 MeV are shown in 
Figs.~\ref{f3}-\ref{f4}. The NF decomposition of the \emph{unsymmetrized} ES amplitude
gives the first Airy minimum A1 located at $\theta\lesssim 100^\circ$ (see Fig.~\ref{f3}), 
which is destroyed by the projectile-target symmetrization (\ref{eq2}). The impact of the 
projectile-target exchange symmetry is shown in Fig.~\ref{f4}, where the NF decomposition 
of the \emph{symmetrized} ES amplitude was done using Eqs.~(\ref{eq16})-(\ref{eq17}). 
While the ES cross section is symmetric about the angle $\theta=90^\circ$, the nearside and 
farside scattering cross sections at angles $\theta<90^\circ$ are symmetrically interchanged 
to the farside and nearside scattering cross sections at angles $\theta>90^\circ$, respectively. 
We note two interesting effects that can be deduced from Fig.~\ref{f4}: 

i) The Airy pattern not destroyed by the symmetrization of the direct and exchange scattering 
amplitudes is reflected symmetrically about $\theta=90^\circ$, and each Airy 
minimum located in the (direct) farside cross section at angle $\theta<90^\circ$ has its 
symmetric partner located in the (exchange) farside cross section at angle $\pi-\theta$. 
For strong refractive systems such as those considered here, the NF cross sections exhibit 
a distinct symmetric Airy minima pattern resembling ``butterfly wings''. 
 
ii) The interference pattern at angles around $90^\circ$ can be interpreted as the interference 
of the nearside and farside components of the total ES amplitude. However, such a NF interference 
of the total ES amplitude is in fact the interference of two farside amplitudes 
(the direct and exchange ones). 

 \begin{figure}[bt]\vspace*{0cm}\hspace*{-0.5cm}
\includegraphics[width=0.55\textwidth]{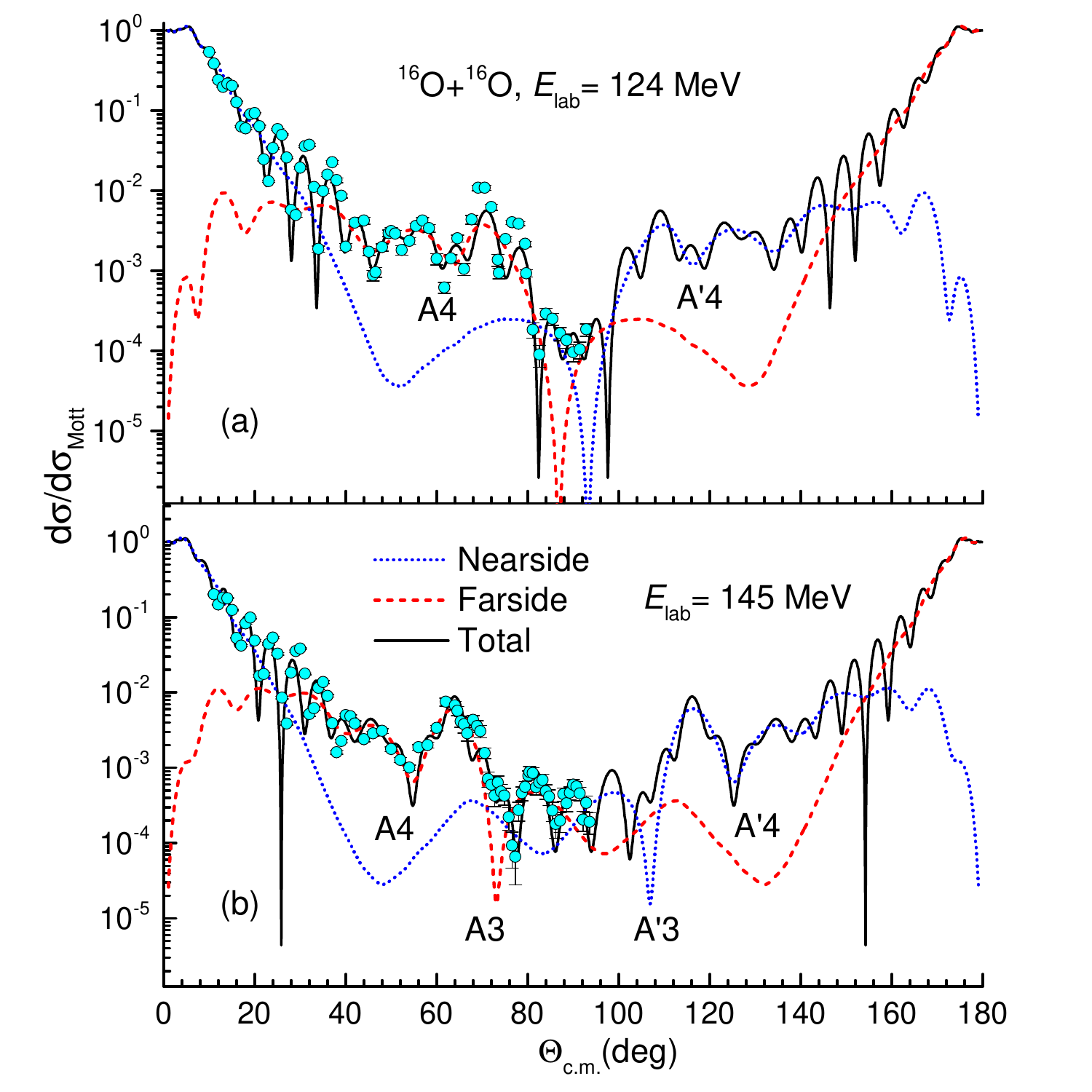}\vspace*{0cm}
\caption{The same as Fig.~\ref{f4} but for the elastic \oo scattering data measured at 
$E_{\rm lab}=124$ MeV \cite{Sugi93} (a), and 145 MeV \cite{Kondo96} (b).} \label{f5}
\end{figure}
The OM results for the elastic \oo scattering at $E_{\rm lab}=124$ and 145 MeV are shown 
in Fig.~\ref{f5}, and one observes the same ``butterfly-wings'' Airy pattern that is symmetric about 
$\theta=90^\circ$ due to the projectile-target exchange symmetry. We note that the refractive, 
nuclear rainbow pattern was well established in elastic \cc and \oo scattering at higher energies, 
like the prominent primary rainbow observed for the \oo system at $E_{\rm lab}=350$ MeV 
\cite{Sti89,Kho95}. However, as energy increases, the broad Airy pattern of the farside cross
section is shifted to smaller angles, and the ES cross section at angles around $90^\circ$ 
merges deeply into the dark side of nuclear rainbow and is, therefore, too small to be measurable. 
Therefore, the low-energy elastic \cc and \oo data \cite{Sto79,Sugi93,Kondo96} are very 
valuable for the study of the projectile-target exchange symmetry of these identical systems.

As pointed out in ii), the oscillation pattern of the elastic cross section around $\theta=90^\circ$ 
is resulted from an interference of two farside amplitudes, the direct and exchange ones, which 
are generated by the same OP. Therefore, the ES cross section at $\theta$ around $90^\circ$ 
is expected to be particularly sensitive to the real OP at sub-surface distances, which would help 
to determine the real OP with less ambiguity  \cite{Kho07r}.
\begin{figure}[bt]\vspace*{-0.5cm}\hspace*{-0.5cm}
\includegraphics[width=0.57\textwidth]{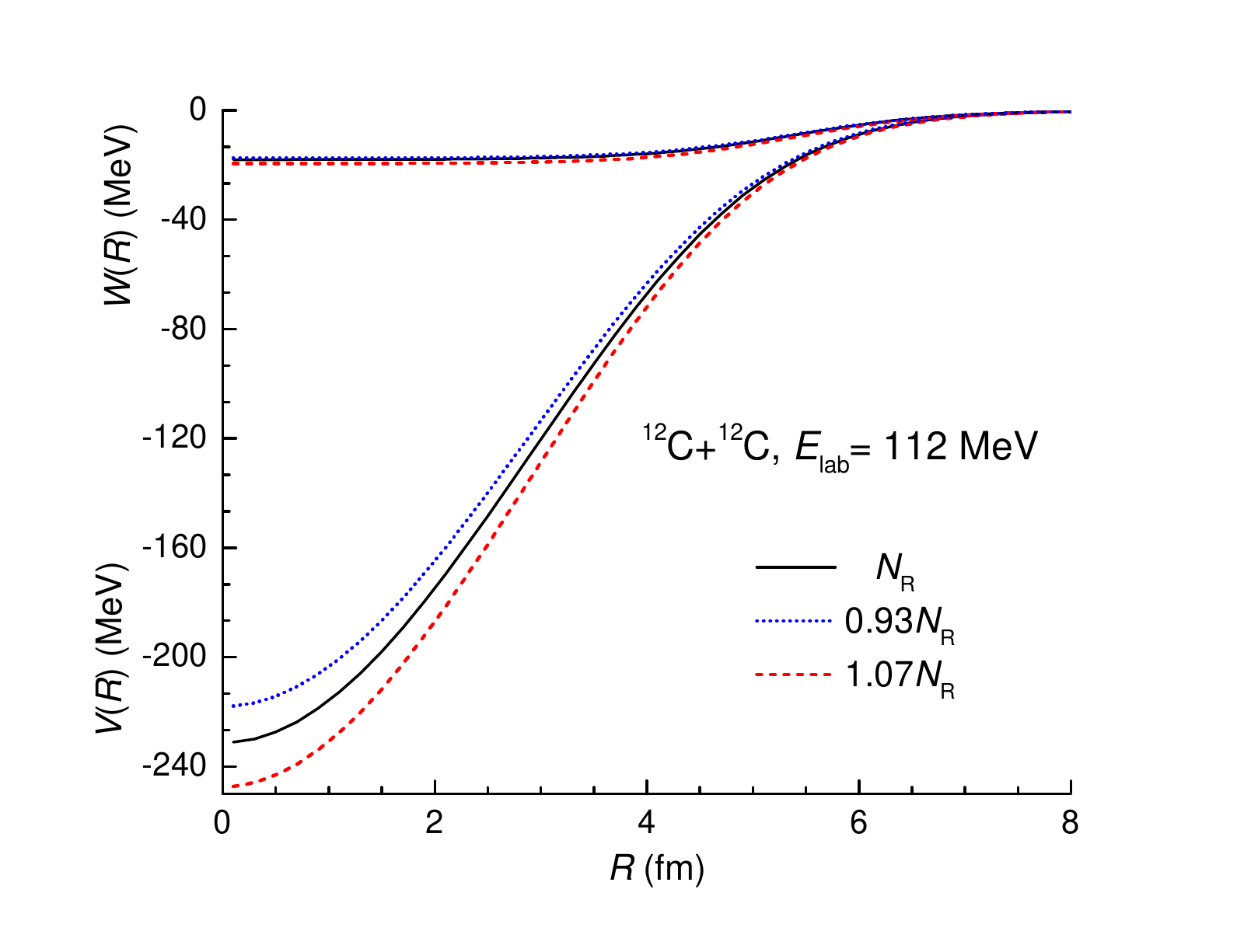}\vspace*{-0.5cm}
\caption{Radial shapes of the real double-folded potential $V$ rescaled up and down by 
around 7\% from that given by the best-fit $N_{\rm R}\approx 1.177$, and the 
corresponding WS imaginary potential of the total OP for the \cc system at 
$E_{\rm lab}=112$ MeV.} \label{f6}
\end{figure}
\begin{figure}[bt]\vspace*{0cm}\hspace*{-0.5cm}
\includegraphics[width=0.55\textwidth]{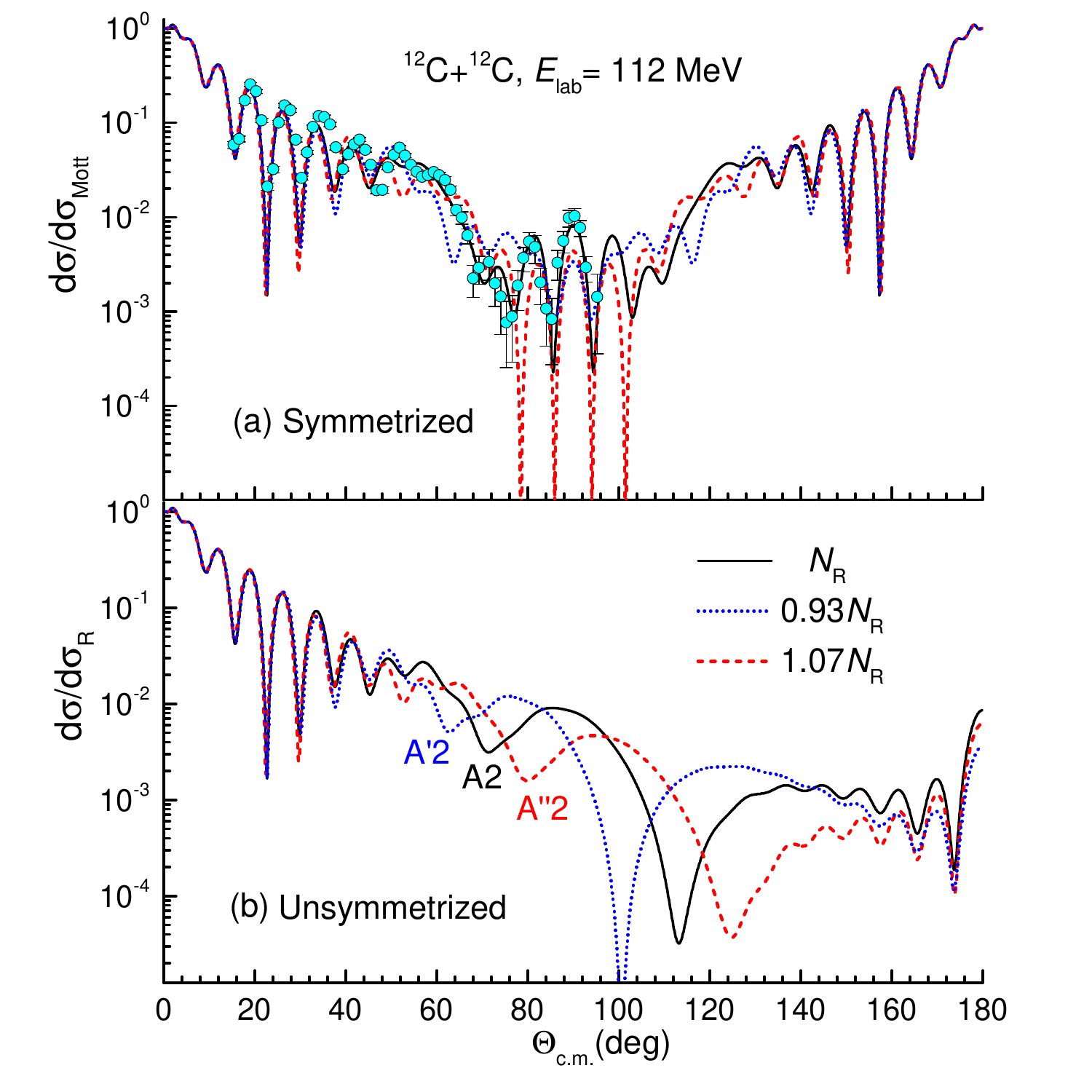}\vspace*{-0.5cm}
\caption{(a) Elastic \cc data measured at $E_{\rm lab}=112$ MeV \cite{Sto79} in comparison 
with the OM results given by the real folded OP rescaled up and down by around 7\% from that
given by the best-fit $N_{\rm R}\approx 1.177$ as shown in Fig.~\ref{f6}. \\
(b) Results of the \emph{unsymmetrized} OM calculation using three choices of the real OP, 
which result in three different locations of the second Airy minimum A2 of the (direct) farside 
cross section.} \label{f7}
\end{figure}

We have explored such a sensitivity to the real OP of the elastic \cc data measured at 
$E_{\rm lab}=112$ MeV by slightly rescaling the strength of the best-fit real folded OP up and 
down by $\approx 7\%$, with the depth of the WS imaginary OP being adjusted in each case by 
$\chi^2$ fit to the measured data. The radial shapes of the complex \cc OP are shown in Fig.~\ref{f6}, 
and the corresponding OM results are compared with the data in Fig.~\ref{f7}.  
It can be seen in panel (b) of Fig.~\ref{f7} that the Airy pattern of the (direct) farside scattering 
cross section is very sensitive to the strength of the real OP. The rescaling of the real OP shown 
in Fig.~\ref{f6} results in different locations of the Airy minima of the direct farside cross 
section, which strongly affect the interference of the direct and exchange farside amplitudes 
at angles around $90^\circ$, as shown in panel (a) of Fig.~\ref{f7}. 
Similar sensitivity of the elastic \cc scattering data at $E_{\rm lab}=78$ MeV \cite{Sto79} 
to the real OP was used in a recent study of the \cc fusion \cite{Chien18} to probe 
different treatments of the nuclear overlap density in the double-folding calculation of the 
real OP of the \cc system at astrophysical energies. 
 
In summary, we have applied the newly developed NF decomposition method for the ES 
amplitude of two identical nuclei to show the effects caused by the projectile-target exchange 
symmetry to the nuclear rainbow pattern in elastic \cc and \oo scattering at low energies.

\section{Elastic \oc scattering and the core-core exchange effect}\label{sec4} 
A similar interchange of the NF scattering as that found above for the identical 
\cc and \oo systems might also be seen in a nonidentical system that has the core-core 
symmetry. We focus here on the \oc system that was considered as a good candidate 
for the observation of nuclear rainbow \cite{Bra91,Bra01}. Several experiments were carried out 
to measure elastic \oc scattering with high precision at low and medium energies, covering 
a wide angular range. We mention here elastic \oc data measured by the Kurchatov group 
\cite{Oglo98,Oglo00,Glu01-181,Glu07} and Strasbourg group \cite{Nico00}.

\begin{figure}[bt]\vspace*{-0.5cm}\hspace*{-0.4cm}
	\includegraphics[width=0.52\textwidth]{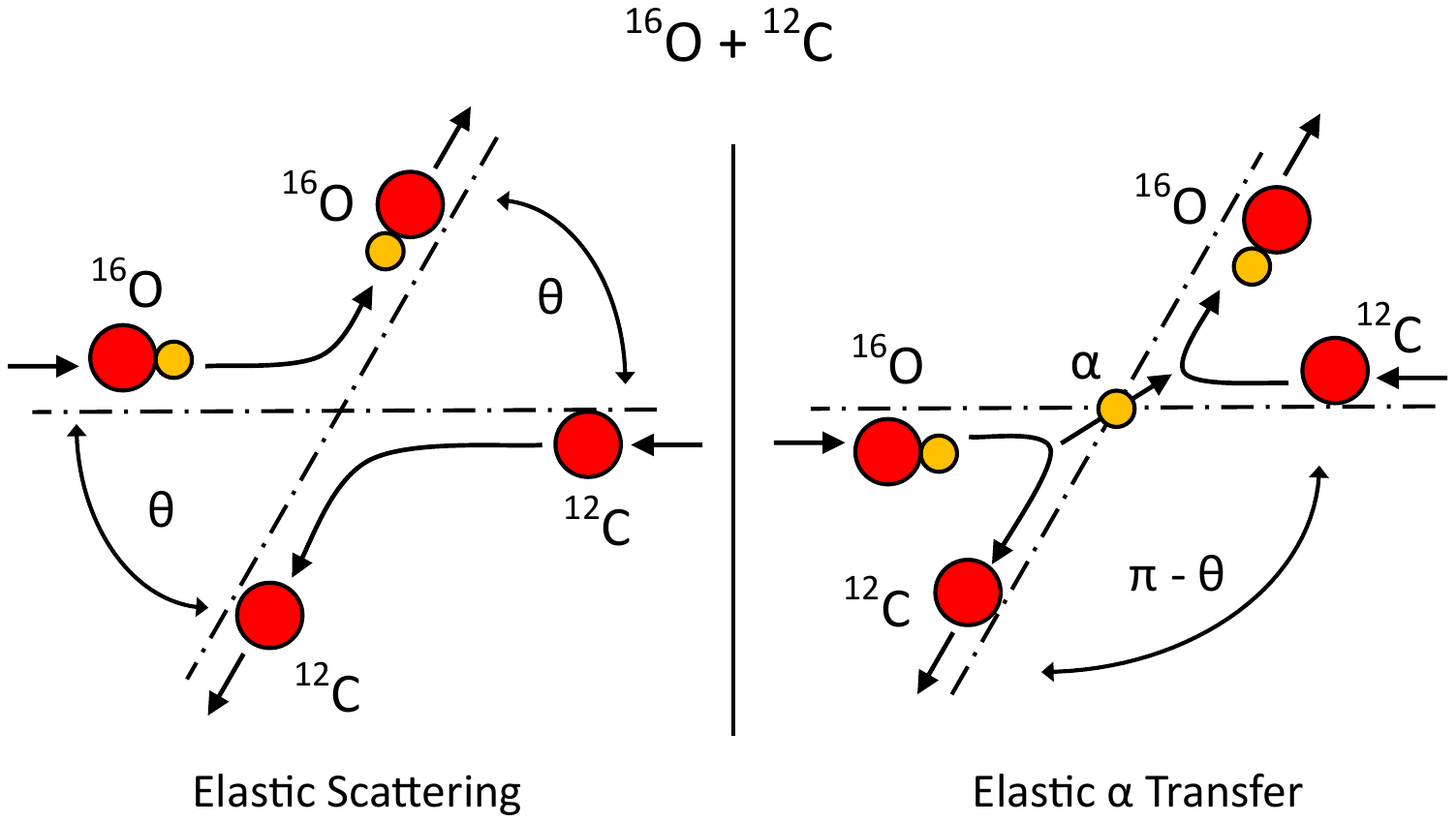}\vspace*{-1.3cm}
	\caption{Kinematical illustration of the elastic scattering and elastic $\alpha$ 
		transfer processes in the \oc system.} \label{f8}
\end{figure}

Different OM studies of elastic \oc scattering (see, e.g., \cite{Kho16,Oglo98,Oglo00}) have shown 
unambiguously the nuclear rainbow pattern of a broad Airy oscillation of the farside cross section. 
However,  at low energies ($E_\text{lab}\lesssim 132$ MeV) the broad pattern
of Airy oscillation is destroyed by a quick oscillation of elastic \oc cross section 
at backward angles. Such a distortion of nuclear rainbow is due mainly to the elastic $\alpha$ 
transfer (ET) between two $^{12}$C cores as illustrated in Fig.~\ref{f8} 
(see, e.g., Refs.~\cite{Szi02,Rud10,Phuc18,Phuc21,Fer19}). 
Recent coupled reaction channel (CRC) studies of elastic \oc scattering, taking into account
the coupling between the ES and ET  channels \cite{Phuc18,Phuc21} have shown that the ET 
process significantly enhances the nearside cross section at backward 
angles, indicating that an interchange of the NF scattering seems to take place in 
elastic \oc scattering. 

\begin{figure}\vspace*{-0.5cm}\hspace*{-0.3cm}
\includegraphics[width=0.54\textwidth]{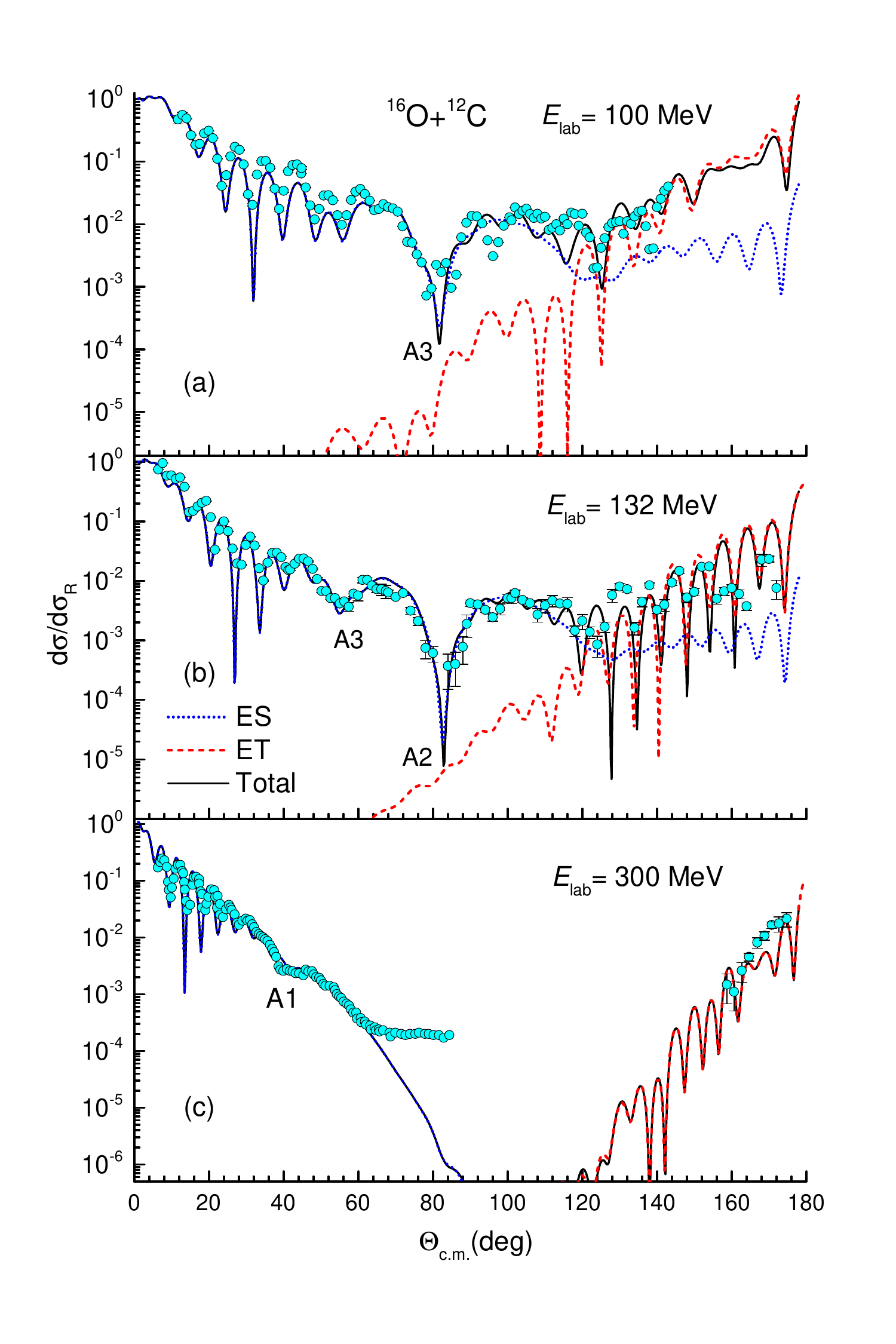}\vspace*{-1cm}
 \caption{Two-channel CRC description of elastic \oc data measured at $E_{\rm lab}=100$ 
MeV (a), 132 MeV (b), and 300 MeV (c) \cite{Nico00,Oglo98,Oglo00,Bra01}, using 
the real double-folded and WS imaginary OP. The ES, ET, and total (ES+ET) elastic 
cross sections are shown as dotted, dashed, and solid lines, respectively. } \label{f9}
\end{figure}

Because of two identical $^{12}$C cores, the ES channel $^{12}$C($^{16}$O,$^{16}$O)$^{12}$C 
and ET channel \Atrans have the same final state which is indistinguishable for the detector 
(see Fig.~\ref{f8}). As a result, the total (nonlocal) wave function of the \oc system is composed 
of both the ES and ET components, with prime indicating the relative coordinate in the ET 
channel \cite{Tho09}
\begin{eqnarray}
\Psi_{\rm total}({\bm r},{\bm r}')&\sim&\left[\Psi_{\rm ES}({\bm r})+\Psi_{\rm ET}(-{\bm r}')\right],
\nonumber\\ 
\mbox{where}\ {\bm r}&=&{\bm r}_1-{\bm r}_2,\ {\bm r}'={\bm r}'_1-{\bm r}'_2. \nonumber
\end{eqnarray}
Like the total ES amplitude of two identical nuclei (\ref{eq14}), the total elastic amplitude 
of the \oc system can be expressed as a coherent sum of the ES amplitude at $\theta$ and ET amplitude 
at $\pi-\theta$ \cite{vOe75,Frahn80s,Frahn80,Phuc19} 
\begin{equation}
 f_{\rm total}(\theta)=f_{\rm R}(\theta)+f(\theta)+f_{\rm ET}(\pi-\theta)
=f_{\rm ES}(\theta)+f_{\rm ET}(\pi-\theta), \label{eq20}
\end{equation}
where $f_{\rm ES}$ and $f_{\rm ET}$ are given by the CRC solutions obtained for the ES 
and ET channels, respectively \cite{Tho88,vOe75}. One can see from Eqs.~(\ref{eq14}) 
and (\ref{eq20})  that $f_{\rm ET}(\pi-\theta)$ is analogous to $f_{\rm EX}(\pi-\theta)$ 
of the identical system. The ET amplitude can be expanded over a partial wave series as 
\begin{equation}
 f_{\rm ET}(\pi-\theta)=\frac{1}{2ik}\sum_\ell(2\ell+1)e^{2i\sigma_\ell}
 (-1)^\ell S_{\rm ET}^{(\ell)} P_\ell(\cos\theta), \label{eq21}
\end{equation}
where $S_{\rm ET}^{(\ell)}$ is the elastic transfer $S_{\rm ET}$-matrix element for the $\ell$-th 
partial wave. The total elastic amplitude (\ref{eq20}) can then be expressed as
\begin{eqnarray}
 f_{\rm total}(\theta)&=&f_{\rm R}(\theta)+\frac{1}{2ik}\sum_\ell(2\ell+1)e^{2i\sigma_\ell}\nonumber\\
 &&\times \left[S_{\rm ES}^{(\ell)}+(-1)^\ell S_{\rm ET}^{(\ell)}\right]P_\ell(\cos\theta). \label{eq22}
\end{eqnarray}
At variance with the total elastic scattering amplitude (\ref{eq9}) of two identical nuclei, 
summation of the partial wave series (\ref{eq22}) is done over both \emph{odd} and \emph{even} 
partial waves $\ell$, with the parity-dependent elastic $\alpha$ transfer $S_{\rm ET}$ matrix added 
to the elastic scattering $S_{\rm ES}$ matrix. The interference of $S_{\rm ES}$ and $S_{\rm ET}$ 
leads to the oscillation of elastic \oc cross section observed at large angles (see Fig.~\ref{f9}). 
Schematically, the elastic $\alpha$ transfer shown in Fig.~\ref{f8} can also be treated as 
the core-core exchange that naturally leads to that interference. 

In this work, the coupling between the ES and ET channels is taken explicitly into account by solving 
the two-channel CRC equations using the code FRESCO \cite{Tho88,Tho09}, to obtain $S_{\rm ES}^{(\ell)}$ 
and $S_{\rm ET}^{(\ell)}$ separately for each partial wave. Although the final state of these two channels 
is the same, there is no way to link explicitly both $S_{\rm ES}$ and $S_{\rm ET}$ to the same \oc scattering 
potential as might naively be expected from a comparison of Figs.~\ref{f2} and \ref{f8}. While $S_{\rm ES}$ 
is associated with elastic (Coulomb+nuclear) scattering, $S_{\rm ET}$ represents the elastic $\alpha$ 
transfer between two $^{12}$C cores which is associated with the dissociation $^{16}$O$\to\alpha+^{12}$C. 
Within the two-channel CRC formalism, $S_{\rm ES}$ and $S_{\rm ET}$ are determined separately using 
the \oc optical potential and the (nonlocal) $\alpha$-transfer interaction potential, respectively 
(we refer to Ref.~\cite{Phuc18} for further details). We note that in the one-channel OM study, one can 
effectively mimic the ET by adding an angular-momentum or parity dependent term to the \oc optical potential 
\cite{vOe75,Phuc19}. In a similar manner, the ET can also be represented by a modified elastic scattering 
$S$ matrix \cite{Frahn80,Frahn80s} that contains an $\ell$-dependent component like that in Eq.~(\ref{eq22}). 
 
The elastic \oc cross sections given by the total elastic amplitude (\ref{eq22}) obtained from solutions 
of the two-channel CRC calculation are compared with elastic \oc data measured 
at $E_{\rm lab}=100$ MeV \cite{Nico00}, 132 MeV  \cite{Oglo98,Oglo00}, and 
300 MeV \cite{Bra01} in Fig.~\ref{f9}, and one can see that the enhanced oscillating 
elastic cross sections at backward angles are mainly caused by the ET or the core-core 
exchange at these energies. Especially, the elastic \oc data measured at 300 MeV at the most 
backward angles are totally due to the ET as can be seen in panel (c) of Fig.~\ref{f9}. 
The good CRC description of elastic \oc data shown in Fig.~\ref{f9} is obtained consistently 
at three energies with the $\alpha$ spectroscopic factor $S_\alpha\approx 1.96$, in agreement 
with the earlier DWBA and two-channel CRC results \cite{Szi02,Mor11}. Although this 
$S_\alpha$ value is larger than those predicted by the shell model (SM) \cite{Volya15} 
or $\alpha$-cluster model \cite{Yama12}, it can be adopted as an effective $S_\alpha$ 
factor for the description of the ET process in the two-channel CRC calculation. 
In fact, a more comprehensive CRC calculation of elastic \oc scattering, coupling up 
to 10 reaction channels of both direct and indirect (multistep) $\alpha$ transfers 
\cite{Phuc18}, accounts well for elastic \oc data at backward angles using $S_\alpha$ 
predicted by the SM calculation \cite{Volya15}. We note here that the (extended) continuum 
discretized coupled channel (CDCC) approach \cite{Doh21}, including the core-core 
exchange into the nonlocal CDCC calculation of low-energy elastic \oc scattering, shows 
the same dominant contribution of the core-core exchange to elastic \oc cross section 
at large angles.  
 
\begin{figure}\vspace*{-0.5cm}\hspace*{-0.3cm}
\includegraphics[width=0.54\textwidth]{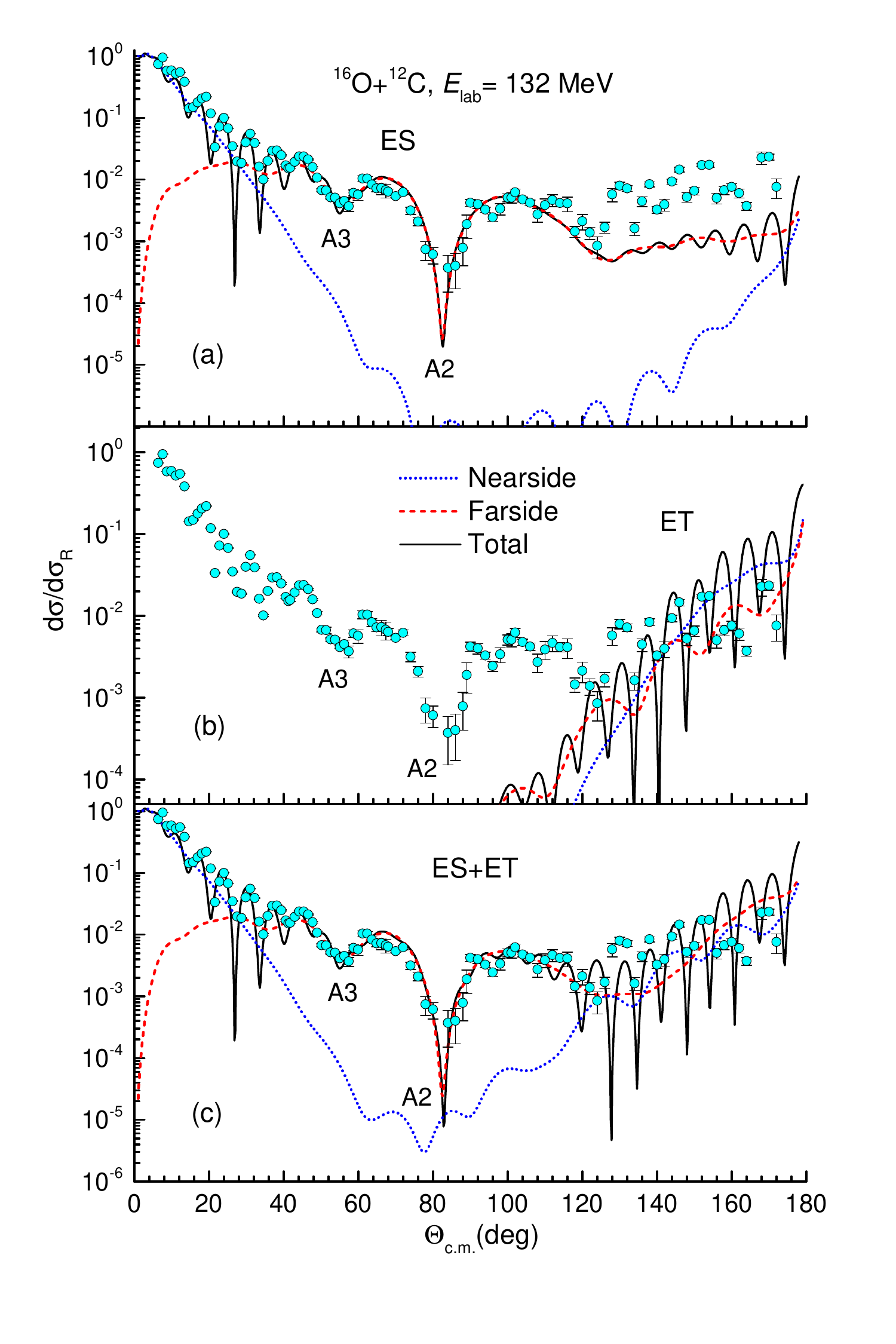}\vspace*{-1cm}
 \caption{NF decomposition of the elastic \oc amplitude at $E_{\rm lab}=132$ MeV given
by the two-channel CRC calculation using the same OP as that used in panel (b) of Fig.~\ref{f9}, 
(a) - for the purely elastic scattering (ES) only, (b) - for the elastic $\alpha$ transfer (ET) only, 
and (c) - for the total (ES+ET) elastic amplitude. } \label{f10}
\end{figure}

Note that the ES and ET contributions to the total elastic \oc cross section are not symmetrically 
equal as the direct and exchange scattering amplitudes found above for two identical nuclei. 
To explore this effect in more details, the NF decomposition of the elastic \oc amplitude, given by 
the two-channel CRC calculation at $E_{\rm lab}=132$ MeV, has been done separately
for the ES, ET, and total (ES+ET) elastic amplitude using Fuller's method \cite{Ful75}. One can see 
in panel (b) of Fig.~\ref{f10} that the ET cross section is \emph{not} a symmetrically reflected pattern 
of the ES cross section shown in panels (a). While the ES cross section is dominated by the farside 
scattering over a wide angular range, the ET cross section is a typical NF interference pattern, 
which indicates a surface character of the ET between two $^{12}$C cores.  This is natural because 
the contribution of $f_{\rm ET}(\pi-\theta)$ to $f_{\rm total}(\theta)$ at the most backward angles 
represents in fact the ET process occurring physically at the most forward angles. 

At variance with the ES occurring at forward angles that includes both the Coulomb and 
nuclear scattering, the repulsive Coulomb interaction does not contribute to the ET process 
which occurs at backward angles. As a result, we found a significant \emph{non-refractive} 
farside component of the ET cross section shown in panel (b) of Fig.~\ref{f10}. Although generated 
by the attractive nuclear interaction, such a farside cross section cannot be associated with nuclear 
rainbow due to a quick modulation of the ET farside cross section caused mainly by the diffraction 
of the $\alpha$-transfer wave. 

Despite the difference discussed above for the ES and ET cross sections, the NF components of the 
total elastic \oc amplitude behave in a manner similar to that shown above for the identical \cc and 
\oo systems. Namely, 
\begin{eqnarray} 
f^{\rm (N)}_{\rm total}(\theta)&=&f^{\rm (N)}_{\rm ES}(\theta)+f^{\rm (F)}_{\rm ET}(\pi-\theta),  
 \label{eq23} \\
f^{\rm (F)}_{\rm total}(\theta)&=& f^{\rm (F)}_{\rm ES}(\theta)+f^{\rm (N)}_{\rm ET}(\pi-\theta).  
 \label{eq24}
\end{eqnarray} 
Like the total ES amplitude of two identical nuclei (\ref{eq16})-(\ref{eq17}), the nearside component 
(\ref{eq23}) of the total elastic \oc amplitude is a superposition of the nearside and farside components 
of the ES and ET amplitudes determined at angles $\theta$ and $\pi-\theta$, respectively, and 
vice versa for the farside component  (\ref{eq24}) of the total elastic amplitude. At the most backward 
angles, the ET process become dominant and the \emph{nearside} component of the total elastic 
amplitude is determined entirely by the \emph{farside} component of the ET amplitude, and vice versa 
for the farside component of the total scattering amplitude, as shown in panels (b) and (c) of Fig.~\ref{f10}. 
Although the NF pattern is not symmetric as found for the elastic scattetring of two identical nuclei, 
the NF interchange between the total elastic and ET amplitudes is naturally caused by the ET between 
two identical $^{12}$C cores, i.e., the core-core exchange symmetry of the \oc system. 

\section{Summary}
The Fuller method of the NF decomposition of the \AA elastic scattering amplitude \cite{Ful75} 
has been generalized for two identical (spin-zero) nuclei, with the projectile-target exchange 
symmetry taken exactly into account. It is shown that the exchange symmetry of two identical 
nuclei results in the symmetric interchange of the nearside and farside scattering cross sections 
at angles passing through $\theta=90^\circ$. 
As a result, the Airy pattern of the nuclear rainbow is reflected symmetrically about $\theta=90^\circ$, 
and each Airy minimum located in the (direct) farside cross section at angle $\theta<90^\circ$ has its 
symmetric partner located in the (exchange) farside cross section at angle $\pi-\theta$. 
 
The Mott interference pattern observed for the identical \cc and \oo systems at medium angles 
was found to be an interference of two farside (direct and exchange) scattering amplitudes, 
which results in an increased sensitivity of elastic scattering cross section around 
$\theta\approx 90^\circ$ to the real OP at sub-surface distances. 
Therefore, elastic \cc and \oo scattering data measured at low energies \cite{Sto79,Sugi93,Kondo96} 
can serve as helpful probes of different theoretical models of the OP for these identical systems. 
Moreover, our extended NF decomposition allows for more accurate NF analyses of identical 
particle scattering at low energies \cite{Hag24}, where nearside and farside amplitudes are 
comparable.

A similar NF interchange of the total elastic and elastic $\alpha$ transfer amplitudes was found in 
the nonidentical \oc system with the core-core symmetry, where the ET process becomes dominant 
at backwards angles. At variance with the pure elastic scattering, a significant \emph{non-refractive} 
farside component of the ET cross section was found, which indicates the surface character of the $\alpha$ 
transfer between two $^{12}$C cores.

\section*{Acknowledgments}
The present research has been supported, 
in part, by the National Foundation for Science and Technology Development of Vietnam
(NAFOSTED Project No. 103.04-2021.74).

\end{document}